# Machine Intelligence Techniques for Next-Generation Context-Aware Wireless Networks

Tadilo Endeshaw Bogale[1,2], Xianbin Wang[1] and Long Bao Le[2]
Western University, London, Canada[1]
Institute National de la Recherche Scientifique (INRS),
Université du Québec, Montréal, Canada[2]
Email: {tadilo.bogale, long.le}@emt.inrs.ca, xianbin.wang@uwo.ca

**Abstract**

The next generation wireless networks (i.e. 5G and beyond), which would be extremely dynamic and complex due to the ultra-dense deployment of heterogeneous networks (HetNets), poses many critical challenges for network planning, operation, management and troubleshooting. At the same time, generation and consumption of wireless data are becoming increasingly distributed with ongoing paradigm shift from people-centric to machine-oriented communications, making the operation of future wireless networks even more complex. In mitigating the complexity of future network operation, new approaches of intelligently utilizing distributed computational resources with improved context-awareness becomes extremely important. In this regard, the emerging fog (edge) computing architecture aiming to distribute computing, storage, control, communication, and networking functions closer to end users, have a great potential for enabling efficient operation of future wireless networks. These promising architectures make the adoption of artificial intelligence (AI) principles which incorporate learning, reasoning and decision-making mechanism, as natural choices for designing a tightly integrated network. Towards this end, this article provides a comprehensive survey on the utilization of AI integrating machine learning, data analytics and natural language processing (NLP) techniques for enhancing the efficiency of wireless network operation. In particular, we provide comprehensive discussion on the utilization of these techniques for efficient data acquisition, knowledge discovery, network planning, operation and management of the next generation wireless networks. A brief case study utilizing the AI techniques for this network has also been provided.

## I. INTRODUCTION

The advent of the fifth generation (5G) wireless network and its convergence with vertical applications constitute the foundation of future connected society which is expected to support 125 billion devices by 2030 (IHS Markit). As these applications and devices are featured by ubiquitous connectivity requirements, future 5G and beyond networks are becoming more complex. Besides the complexity increase of base stations (BSs) and user equipments (UEs), significant challenges arise from the initial network planning to the deployment and situation-dependent operation and management stages.

The network architecture of 5G and beyond will be inevitably heterogeneous and multi-tier with ultra-dense deployment of small cells to achieve the anticipated 1000 fold capacity increase cost-effectively. For instance, the mixed use of planned and centrally controlled macro-BSs and randomly deployed wireless fidelity (WiFi) access points or femto-BSs in the ultra-dense heterogeneous network (HetNet) raises several unexpected operation scenarios, which are not possible to envision at the network design stage. This requires future wireless networks to have self organizing, configuring and healing capability based on the operational condition through tight coordination among different nodes, tiers and communication layers. These challenges make the existing network design strategies utilizing a fairly simple statistics experience unacceptable performance (for example, in terms of spectrum

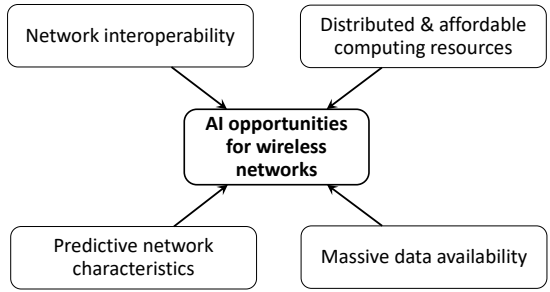

Fig. 1. Favorable conditions for the adoption of machine intelligence techniques in the next generation wireless networks.

and energy efficiency, coverage, delay and cost) [1], [2].

The rapidly growing number of machine-type-communication (MTC) devices contribute a considerable portion on the complexity of this ultra-dense network. Many of the future MTC applications supported by 5G and beyond will require the underlying wireless networks to achieve high availability, reliability and security, very short transit time and low latency [3]. Furthermore, in such use cases, uninterrupted and safe operation is often the top priority (for instance, connected vehicles). Taking a MTC application offline for any reason can cause significant business loss or non-tolerable customer experience, and many of the MTC devices are resource-constrained and will not be able to rely solely on their own limited resources to fulfill their processing demands [4].

As a result, these latency critical applications cannot be moved to the network controller or cloud due to delay, bandwidth, or other constraints. Moreover, the data sets generated from these devices will be extremely diverse and may have large-scale missing (inaccurate) values [5]. In addition, a number of new data hungry MTC immersive use-cases will arise including wearables, virtual realities, intelligent product and support-systems where most of them will use built-in back-end data infrastructure and analytics engine to provide context-aware services. All these necessitates the next generation network (i.e. 5G and beyond) to adopt an intelligent and context-aware approach for network planning, design, analysis, and optimization.

We are in the beginning phase of an intelligent era that has been driven by the rapid evolution of semiconductor industries, computing technologies, and diverse use cases. This is witnessed by the tight integration of networked information systems, sensing and communication devices, data sources, decision making, and cyber physical infrastructures. The proliferation of tiny wireless sensors and MTC devices, and smart phones also show clear evidences for the exceptional processing capability and cost-effectiveness of semiconductor devices. These promising developments facilitate distributed computing resources not only in the cloud but also in the fog and edge nodes. Both fog and edge computing attempt pushing the intelligence and processing capabilities down closer to where the data originates.

The edge computing aims to integrate intelligence and processing power capabilities closest to the original data source. The edge node, for example, intelligent programmable automation controllers (PACs), determines which raw input data should be stored locally or sent to the fog (cloud) for further analysis. On the other hand, in the fog computing, all the raw input data will first be converted to the appropriate Internet protocol (such as HTTP) before sending to the fog nodes. Thus, higher-level data contents are processed, stored and sent to the cloud for further analysis in the fog devices (for example, intelligent routers, access points, Internet of things (IoT) gateways). Thus, the edge and fog enabled network allows distributed computing, storage, control, communication and networking functions by reducing the data transmitted and work load of the cloud, latency and system response time especially for applications demanding localized and location dependent information [6]. Moreover, the node, user, sensor, or MTC device is potentially capable of generating row and processed data at different granularity levels which ultimately help the network to have a massive amount of data likely exhibiting some pattern. This will help different nodes to leverage data mining and analytics techniques to predict relevant network metrics such as user mobility, traffic behavior, network load fluctuation, channel variations, and interference levels.

All these opportunities enable efficient and flexible resource allocation and management, protocol stack configuration, and signaling procedure and physical layer optimization, and facilitate existing devices to harness the powers of sensors, edge, fog and cloud based computing platforms, and data analytics engines [7]–[9]. These also create favorable conditions to engineer a tightly integrated wireless network by adopting the AI principles (see Fig. 1) incorporating learning, reasoning and decision making mechanisms which are crucial to realize context-awareness capability. A typical next generation network utilizing the AI principles at different nodes is shown in Fig. 2. Towards this end, the current paper provides a comprehensive survey on the utilization of AI integrating machine learning, data analytics and natural language processing (NLP) techniques for enhancing the efficiency of wireless systems. We particularly focus on the utilization of these techniques for efficient wireless data acquisition and knowledge discovery, planning, and operation and management of the next generation wireless networks. A brief case study showing the utilization of AI techniques for this network has also been provided.

The paper is organized as follows. In Section II, we discuss data acquisition and knowledge discovery approaches used in AI enabled wireless. Then, a comprehensive discussion on how these knowledge can be used in network planning, operation and management of the next generation wireless is given in Sections III and IV. A case study discussing the applications of AI techniques for channel impulse response (CIR) prediction and context-aware data transmission is then provided in Section V. Finally, conclusions are drawn in Section VI.

## II. Data Acquisition and Knowledge Discovery

Efficient data acquisition and knowledge discovery is one of the requirements of future wireless networks as it helps to realize situation aware and optimized decisions as shown in Fig. 3. The gathered data may need to be processed efficiently to extract relevant knowledge. Furthermore, as the available data may contain a large amount of erroneous (missing) values, a robust knowledge discovery may need to be devised [5].

### A. Data Acquisition

The AI based tools relying on machine learning for the input data mining and knowledge model extraction at different levels could be applied [10]. This includes the cell level, cell cluster level and user level. In general, one can collect data from three main sources; network, user, and external. The network data characterizes the network behavior including outage and usage statistics of services or nodes, and load of a cell. The user data could comprise user subscription information and user device type. And, the external data contains a user specific information obtained from different sources such as sensors and channel measurements [11].

One way of collecting wireless data is by employing content caching where the idea is to store popular contents at the network edge (at BSs, devices, or other intermediate locations). In this regard, one can enable the proactive cache type if the traffic learning algorithm predicts that the same content will be requested in the near future [12], [13]. Moreover, since different users may request the same content with different qualities, each edge node may need to cache the same content in different granularity (for example, caching video data with different resolution). This further requires the edge device to apply coded (adaptive) caching technique based on the quality of service (QoS) requirement of the requester [14]. Coded caching also enables devices to create multicasting opportunities for certain contents via coded multicast transmissions [15].

In some cases, a given edge (fog) may collect date from more than one sources with different connectivity criteria [16]. In a fog enabled wireless network, this is facilitated by employing IoT devices which leverage a multitude of radio-access technologies such as wireless local area network (WLAN) and cellular networks. In this regard, context-aware data collection from multiple sources probably in a compressed format by employing appropriate dimensionality reduction techniques under imperfect statistical knowledge of the data while simultaneously optimizing multiple objective functions such as delay and transmission power can be enabled [13].

### B. Knowledge Discovery

Efficient knowledge discovery is critical for optimized operation and management of the network. The network may need to use a novel learning technique such as deep learning to extract the hidden contextual information of the network which are crucial for knowledge base (KB) creation. In general, context is related to any information used to characterize the situation of an entity, including surrounding location, identity, preferences, and activities. Context may affect the operation and management procedures of complex systems at various levels, from the physical device level to the communication level, up to the application level [17]. For instance, uncovering the relation between the device and network information (user location, velocity, battery level, and other medium access control (MAC) and higher layer aspects) would permit adaptive communication and processing capa-bilities based on the changes in the environment and application [2], [17].

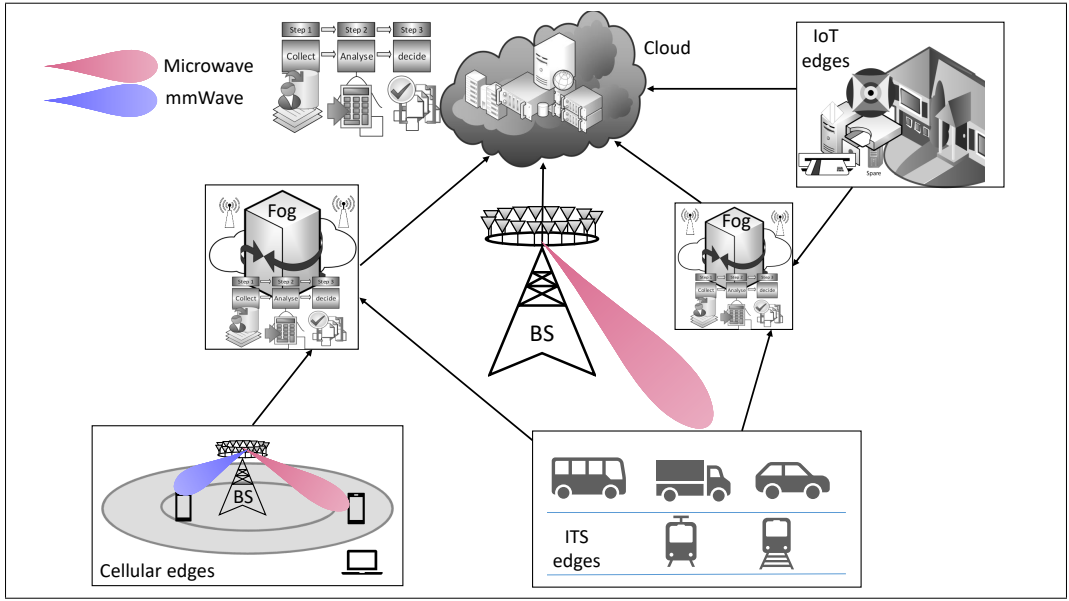

Fig. 2. Typical next generation network adopting AI principles with learning, reasoning and decision making.

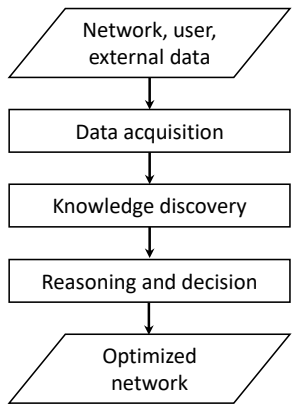

Fig. 3. Optimized network design with AI techniques.

Analyzing wireless data contextually (semantically) facilitate wireless operators to optimize their network traffic. To realize semantic aware traffic optimization, however, the network may need to understand the content of the signal. One way of understanding the information content of a data is by creating semantic aware ontology using predefined vocabulary of terms and concepts [18]. The ontology specification can provide an expressive language with many logical constructs to define classes, properties and their relationships. In this regard, the authors of [18] propose a semantic open data model for sensor data called *MyOntoSens* and write using ontology web language 2 description logic language. The proposed KB has been implemented using *protégé* and pre-validated with pellet reasoner. In a similar context, an ontology for wireless sensor networks (WSNs) dedicated to the description of sensor features and observation has been presented in [19], [20].

Understanding the context also helps to produce a context-aware compressed (summary) information which will utilize less radio resource for transmission. For instance, if a BS would like to transmit text information to a user, the BS can transmit only its contextual coded data. The user will, then, extract the desired content just from the context by utilizing appropriate decoder and big-data analytics technique such as NLP. As context differs from the user's world knowledge about the content, the coding and decoding technique may vary among users [17]. In general, two types of content summarizing approaches are commonly adopted; abstractive and extractive. The extractive approach uses only the relevant content from the original information whereas, the abstractive approach may use new words (expressions or contents) as part of the summary information [21], [22]. Although most of the existing methods can extract useful information for the summary, they are very far from generating human understandable summary. One of the main reasons is the loose associations and unordered information distribution which make it hard to extract syntactically correct and semantically coherence information from the summary. In fact, modeling the coherence of information summary is one of the active areas of research [21].

## III. NETWORK PLANNING

One of the most critical aspects determining the performance of a wireless network is the initial planning. This includes infrastructure (node deployment), frequency, number of parameters and their configuration setting procedures, and energy consumption to run the network in both idle (no data communication takes place between the network and a user) and active (data communication takes place between the network and a user) conditions. A well planned network may need to guarantee satisfactory user QoS (in terms of data rate, reliability and latency) and the network operator requirements (in terms of cost). The machine learning technique can be used for planning different parts of the network by utilizing the network and user data.

### A. Node Deployment, Energy Consumption and RF Planning

The future generation wireless networks will be extremely dense and heterogeneous, equipped likely with moving and flying BSs, and featured by continually varying network conditions [23]. This fact makes the existing network planning techniques, which are mainly static and designed from expensive field tests, not suitable for the future wireless networks [2], [24]. The utilization of AI techniques for network planning has recently received an interest in the research community. For instance, a machine learning technique is suggested for content popularity and request distribution predictions of a cache where its design in the network considers several factors including cache placement and update strategy which are determined by utilizing the users' content request distribution and frequency, and mobility pattern [25].

In [26], an AI based system which leverages graph theory based problem formulations for the fiber to home network is proposed to automate the planning process. To solve the problems, mixed integer linear programming (MILP), ant colony optimization (ACO) and genetic algorithm (GA) have been applied. The authors of [10] employ the principles of AI for radio access network (RAN) planning which includes new cell, radio frequency (RF) and spectrum of the 5G wireless. The analysis is performed by processing input data from multiple sources, through learning based classification, prediction and clustering models, and extracting relevant knowledge to drive the decisions made by 5G network.

Wireless networks contribute an increasing share in the energy consumption of the information communications technology (ICT) infrastructure. Over 80% of a wireless network power consumption is used by the RANs, especially at the BS, since the present BS deployment is designed on the basis of peak traffic loads and generally stays active irrespective of the huge variations in traffic load [27]. This makes the current energy planning inefficient for the future smart cities aiming to realize green communication. To enable energy efficient wireless, different AI techniques have been suggested. For instance, the authors of [28] propose a method to realize the efficient use of electricity by autonomously controlling network equipments such as servers, air-conditioners in an integrated manner. Furthermore, the authors of [29] suggest predictive models, including neural network and Markov decision, for the energy consumption of IoT in smart cities. Along this line, a BS switching solution for traffic aware greener cellular networks using AI techniques has also been discussed in [27].

### B. Configuration Parameter and Service Planning

The number of configurable parameters in cellular networks fairly increases when moving from one generation to the next. For instance, in a typical 3G and 4G nodes these parameters are around 1000 and 1500, respectively. It is anticipated that this trend will continue and the recently suggested 5G network node will likely have 2000 or more parameters. In addition, unlike the current and previous generation networks which provides static services, the next generation network may need to support the continuously evolving new services and use cases, and establish sufficient network resource and provisioning mechanisms while ensuring agility and robustness. These necessitate the next generation network to understand parameter variations, learn uncertainties, configure network parameters, forecast immediate and future challenges, and provide timely solutions by interacting with the environment [27]. In this direction, the utilization of big data analytics has been discussed for protocol stack configuration, signaling procedure and physical layer procedure optimizations in [9].

Future smart cities require well planned wired and wireless networks with ubiquitous broadband connectivity, and flexible, real time and distributed data processing capability. Although most modern cities have multiple cellular networks that provide adequate coverage and data processing capability, these networks often have limited capacity and peak bandwidths and fail to meet the real time constraint of different emerging tactile applications. These make the realization of advanced delay critical municipal services envisioned in a smart city (e.g., real-time surveillance, public safety, on-time advisories, and smart buildings) challenging [1]. One way of addressing this challenge would be by deploying an AI integrated fog based wireless architecture which allows data processing of the network using a number of distributed nodes. This will help for analyzing network status, detecting anticipated faults and planning new node deployment using AI techniques [1].

## IV. Network Operation and Management

Energy and spectrum efficiency, latency, reliability, and security are the key parameters that are taken into account during the network operation stage. And properly optimizing these parameters usually yield satisfactory performance for both the service providers and end users. In addition, these optimization parameters usually require simple and real time learning and decision making algorithms.

### A. Resource Allocation and Management

Different AI techniques have been proposed for resource allocation, management and optimization of wireless networks such as cellular, wearable, WSN, body area network (BAN) [24]. In [30], the potential of AI in solving the channel allocation problem in wireless communication is considered. It is demonstrated that the AI based approach has shown better performance than those of randomized-based heuristic and genetic algorithms (GAs). In [31], radio access technology (RAT) selection utilizing the Hopfield neural networks as a decision making tool while leveraging the ability of AI reasoning and the use of multi-parameter decision by exploiting the options of IEEE 802.21 protocol is proposed. A machine learning based techniques including supervised, unsupervised, and reinforcement learning techniques, have been exploited to manage the packet routing in many different network scenarios [32]. Specifically, in [33], [34], a deep learning approach for shortest traffic route identification to reduce network congestion is presented. A deep learning technique aiming to shift the computing needs from rule-based route computation to machine learning based route estimation for high throughput packet processing is proposed in [32]. Along this line, a fog computing based radio-access network which exploits the advantage of local radio signal processing, cooperative radio resource management, and distributed storage capability of fog has been suggested to decrease the load on the front haul and avoid large scale radio signal processing in the centralized baseband controllers [1].

The utilization of unlicensed spectrum as a complement to licensed one receives an interest to offload network traffic through the carrier aggregation framework, while critical control signaling, mobility, voice and control data will always be transmitted on the licensed bands. In this aspect, the authors of [35] propose a hopfield neural network scheme for multi-radio packet scheduling. The problem of resource allocation with uplink-downlink decoupling in a long term evolution-unlicensed (LTE-U) system has been investigated in [36] for which the authors propose a decentralized scheme based on neural network. The authors in [37] propose a distributed approach based on Q-learning for the problem of channel selection in an LTE-U system. Furthermore, in a multi-RAT scenario, machine learning techniques can allow the smart use of different RATs wherein a BS can learn when to transmit on each type of frequency band based on the network conditions. For instance, one can apply a machine learning to predict the availability of a line of sight (LoS) link, by considering the users' mobility pattern and antenna tilt, thus allowing the transmission over the millimeter wave band.

### B. Security and Privacy Protection

The inherent shared nature of radio propagation environment makes wireless transmissions vulnerable to malicious attacks, including eavesdropping and jamming. For this reason, security and privacy protection are fundamental concerns of today's wireless communication system. Wireless networks generally adopt separate security level at different layers of the communication protocol stack. Furthermore, different applications usually require different encryption methods [42]. The utilization of AI techniques for wireless security has received a significant interest.

TABLE I
MAIN ISSUES IN AI-ENABLED WIRELESS NETWORK

| Data acquisition and knowledge discovery | Ref. |
|---|---|
| • Context-aware data acquisition from single/multiple sources | [13] |
| • Coded (adaptive) caching | [14], [15] |
| • Semantic-aware Ontology (KB) creation from network data | [18]–[20] |
| • Robust knowledge discovery from erroneous (missing) data | [5] |
| Network planning | |
| • Node deployment and radio frequency allocation | [10], [26] |
| • Caching and computing placement and content update | [38] |
| • Energy consumption modeling and prediction (idle/active) | [27], [28] |
| • Parameter and service configuration procedure | [1], [9] |
| Network operation and management | |
| • Resource allocation: RAT and channel selection, packet routing, distributed storage and processing, multi-RAT packet scheduling | [30], [31] [1], [32] [35] |
| • Security: Spoofing attack and intrusion detection | [39]–[41] |
| • Latency: Context-aware edge computing and scheduling | [5], [38] |

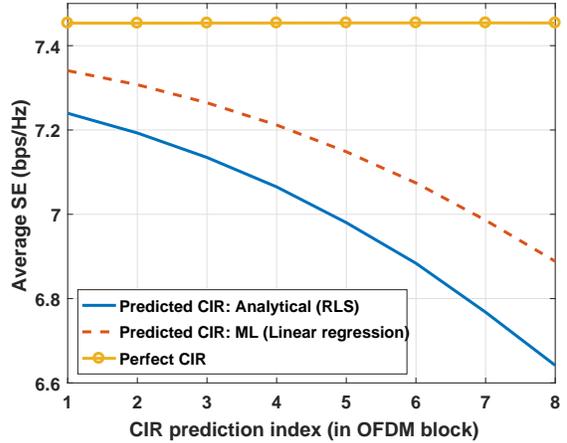

Fig. 4. Comparison of analytical and machine learning (ML) approaches in terms of achieved average SE for different future OFDM block index CIR.

In [39], a spoofing attack detection scheme using random key distribution based artificial immune system (AIS) has been proposed. In a similar path, an approach based on GA and AIS, called GAAIS, for dynamic intrusion detection in mobile ad-hoc network (MANETs) is suggested in [41]. In [40], advanced detection of intrusions on sensor networks (ADIOS) based intrusion detection and prevention system is developed. The ADIOS is designed to mitigate denial-of-service attacks in wireless sensor networks by capturing and analyzing network events using AI and an expert system developed using the C language integrated production system tool. In a similar work, the authors of [43] propose AI based scheme to secure the communication protocol of connected vehicles.

## C. Latency Optimization for Tactile Applications

The next generation wireless networks are featured by several mission critical (tactile) applications such as lane switching in automated vehicles. For vehicular networks, different levels of automation has been defined by the USA department of transportation (DOT) ranging from a simple driver assistance (level 1) to the full automation mode (level 5). For this application, one can apply different message representations including warning alarm, picture and audio information to request intervening. In fact, recently it is demonstrated via experiment that the use of natural language generation techniques from imprecise data improves the human decision making accuracy. Such a linguistic description of data could be designed by modeling vague expressions such as *small* and *large*, which are norms in daily life conversation, using fuzzy logic theory [44]. All these facilitates the utilization of predictive machine learning as in [45], [46].

From the computing side, edge devices can be used for effective low-latency computations, using the emerging paradigm of mobile edge computing. However, optimizing mobile edge computing faces many challenges such as computing placement, computational resource allocation, computing task assignment, end-to-end latency, and energy consumption. In this aspect, a machine learning technique can be used to address these challenges by utilizing historical data. Predicting computational requirements enable the network devices to schedule the computational resources in advance to minimize the global latency. In this aspect, the authors of [38] propose a cross-system learning framework in order to optimize the long-term performance of multi-mode BSs, by steering delay-tolerant traffic towards WiFi. Furthermore, in a fog enabled wireless system, latency can be addressed by exploiting different levels of awareness at each edge network. In fact, a number of learning techniques can be applied to achieve these awareness including incremental, divide and conquer, parallel and hierarchical [5]. A brief summary of different issues in the AI-enabled wireless network is presented in Table I.

## V. DESIGN CASE STUDIES

This section discusses typical design case studies in which the AI techniques can be applied for the context-aware wireless network.

## A. Machine Learning for CIR Prediction

This study demonstrates the utilization of machine learning tools for optimizing wireless system resources. In this aspect, we select the wireless CIR prediction as the design objective. To solve this design objective, the first possibility could be to apply different analytical CIR prediction techniques (for example, the recursive least square (RLS) prediction proposed in [45]). The second possibility could be to predict future CIRs by leveraging the past experience. The former possibility is very expensive particularly when real time prediction is needed. Furthermore, in most cases, the analytical prediction approach may fail whenever there is a modeling error or uncertainty. The latter possibility, however, is simple as it employs the past experience and applies standard vector multiplication and addition operations [47]. This simulation compares the performances of RLS and machine learning prediction approaches. For the machine learning, we employ the well known multivariate linear regression.

For the comparison, we consider an orthogonal frequency domain multiplexing (OFDM) transmission scheme where a BS equipped with $N$ antennas is serving a single antenna IoT device. The CIR is modeled by considering a typical scenario of the IEEE 802.11 standard with channel correlation both spatially and temporarily. The spatial channel covariance matrix is modeled by considering the uniform linear array (ULA) structure, and the temporal channel correlation is designed by the well known Jake's model [48]. The number of multipath taps $L = 4$, fast Fourier transform (FFT) size $M = 64$, $N = 8$, OFDM symbol period $T_s = 166\mu$s, RLS window size $S_b = 8$, forward prediction window size $S_f = 8$, carrier frequency $5.6$GHz, and mobility speed of the IoT device $30$km/hr. The signal to noise ratio (SNR) for each sub-carrier is set to $10$dB. With these settings, Fig. 4 shows the average spectrum efficiency (SE) obtained by the RLS and machine learning approaches for sub-carrier $s = 4$. In both cases, the achieved SE decreases as the future OFDM-block index increases. This is expected since the number of unknown CIR coefficients increase as the future block index increases leading to a degraded CIR prediction quality. However, for a fixed future prediction index, the machine learning approach yields better performance than the RLS one[1].

[1] Note that similar average performance is observed for other sub-carriers.

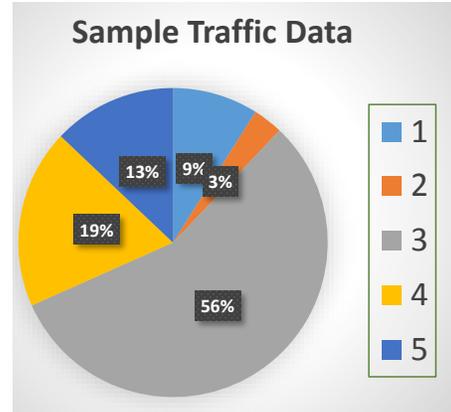

Fig. 5. Data traffic of a sample abstract information: In this figure, (Sets) 1 to 5 denote, Background, Objective, Method and Result, Conclusion, and Related and Future Works, respectively.

## B. Context-Aware Data Transmission using NLP Techniques

Context (semantic) aware information transmission is crucial in the future generation network. To validate this, we employ abstract texts from scientific articles [49]. According to this paper, each scientific abstract text consists of different types of information including research background, methodology, main results etc. Fig. 5 shows the expert annotated data size for different types for 2000 biomedical article abstracts. As can be seen from this figure, different information types use different portions of the overall data set. And for a given user, one can transmit the desired information according to the context. For instance, for a user who is interested in the basics about the article, transmitting the background information could be sufficient which accounts for only 9% of the total traffic. This shows that semantically enabled data transmission will reduce the network traffic while simultaneously maintaining the desired QoS experience of users.

Such a transmission, however, is realized when sentences of similar types are clustered correctly for each abstract. In scientific papers, the location and part-of-speech voice of a sentence are crucial features to identify its class set [49]. We have employed these features with the commonly used data clustering algorithms (i.e., K-means and Agglomerative) and present the accuracy achieved by these algorithms for each type as shown in Table II. As can be seen from this table, different clustering algorithms yield different accuracy. One can also notice from this table that significant research work may need to be undertaken

TABLE II
ACCURACY OF DIFFERENT CLUSTERING METHODS

| Clustering Method | Set 1 | Set 2 | Set 3 | Set 4 | Set 5 |
|---|---|---|---|---|---|
| K-Means | 0.34 | 0.17 | 0.35 | 0.31 | 0.16 |
| Agglomerative | 0.21 | 0.18 | 0.38 | 0.30 | 0.15 |

to reach the ideal performance.

## VI. CONCLUSIONS

The next generation wireless networks, which would be more dynamic, complex with dense deployment of BSs of different types and access technologies, poses many design challenges for network planning, management and troubleshooting procedures. Nevertheless, wireless data can be generated from different sources including networked information systems, and sensing and communication devices. Furthermore, the emerging fog computing architecture aiming for distributed computing, storage, control, communication, and networking functions closer to end users contribute for the efficient realization of wireless systems. This article provides a comprehensive survey on the utilization of AI integrating machine learning, data analytics and NLP techniques for enhancing the efficiency of wireless networks. We have given a comprehensive discussion on the utilization of these techniques for efficient wireless data acquisition and knowledge discovery, planning, operation and management of the next generation wireless networks. A brief case study showing the utilization of AI techniques has also been provided.